# Color-Tunable Mixed-Cation Perovskite Single Photon Emitters

Marianna D'Amato,[1†] Qi Ying Tan,[2,3†] Quentin Glorieux,[1]

Alberto Bramati,[1*] Cesare Soci[2,4*]

[1] Laboratoire Kastler Brossel, Sorbonne Universite, CNRS, ENS-PSL Research University, College de France, 4 place Jussieu, 75252 Paris Cedex 05, France

[2] Centre for Disruptive Photonic Technologies, The Photonics Institute, 21 Nanyang Link, Nanyang Technological University, Singapore 637371

[3] Interdisciplinary Graduate School, Energy Research Institute @NTU (ERI@N), Nanyang Technological University, 50 Nanyang Drive, Singapore 637553

[4] Division of Physics and Applied Physics, 21 Nanyang Link, School of Physical and Mathematical Sciences, Nanyang Technological University, Singapore 637371

*Correspondence to: csoci@ntu.edu.sg, alberto.bramati@lkb.umpc.fr

†These authors, alphabetically ordered, contributed equally to this work.

**Quantum photonics technologies like wavelength division multiplexing (WDM) for high-rate quantum key distribution require narrowband, spectrally tunable single photon emitters. Physical methods that rely on the application of large mechanical strain to epitaxial quantum dots or electric and magnetic fields to color centers in 2D metal dichalcogenides provide limited spectral tunability. Here we adopt a chemical approach to synthesize a family of colloidal mixed-cation perovskite quantum dots ($Cs_{1-x}FA_xPbBr_3$) that show highly photo-stable, compositionally tunable single photon emission at room temperature – spanning more than 30 nm in the visible wavelength spectral range. We find that, tailoring the stoichiometry of the organic formamidinium (FA) cation in all-inorganic cesium lead bromide ($CsPbBr_3$) perovskite quantum dots detunes the electronic band structure while preserving their excellent single photon emission characteristics. We argue that the mixed-cation perovskite quantum dots studied in this work offer a new platform for the realization of color-tunable single photon emitters that could be readily integrated in a diversity of quantum photonic devices.**

**Introduction**

The development of solid-state single photon emitters such as color centers in 2D transition metal dichalcogenides and diamond or III-V and halide perovskite colloidal quantum dots is paving the way to quantum photonic technologies operating at room temperature. Spectral tunability of such single photon emitters is highly sought for advanced applications such as wavelength division multiplexing (WDM) for quantum key distribution. Conventional methods to achieve color-tunability of single photon emitters rely on the application of external perturbations like mechanical strain, electric, and magnetic fields. However, these techniques are hardly scalable and yield small spectral shifts even for large, applied strain and electric or magnetic fields.[1-3] Therefore, alternative color-tunable single photon systems with broader spectral tunability are highly sought.

In recent years, lead halide perovskites have attracted wide interest for their outstanding optoelectronic properties. Owing to their facile solution-processability, compositional tunability, high absorption and photoluminescence efficiency, and excellent charge transport characteristics,[4-8] perovskites have proven to be an outstanding materials platform for solar cells, light emitting devices, and laser applications.[4,9-12] In the last decade, considerable efforts were made to extend the use of halide perovskites to the domain of quantum optics. Single cation lead halide perovskite quantum dots were proven to be a low cost and scalable platform for single photon, coherent quantum emission in the visible.[13-17] Recently, their integration into various scalable solid-state platforms was also successfully demonstrated, confirming their potential for integrated quantum devices.[18]

Spectral tunability of perovskite quantum dots can be easily achieved by tuning their size and composition, without the need to apply mechanical strain or external fields. Notably, the addition of a second A-site cation additive into the single cation perovskites allows fine-tuning the emission wavelengths by controlling the relative cation composition.[19-22] This concept was successfully employed in photovoltaic and light emitting devices, yielding promising efficiency and brightness.[22, 23] Alongside with color-tunability, mixed-cation perovskites bring about an enhancement in structural stability which further translates into better emission stability. As such, mixed-cation perovskites are undertaken as a common approach to overcome the "perovskite red wall", that is the difficulty to obtain stable red and near-infrared photoluminescence (PL).[24] Specifically, the addition of FA cation additives into $CsPbI_3$ shows an enhancement in emission stability compared to $CsPbI_3$.

By extending the compositionally tunable mixed cation perovskite approach down to the single photon level, here we look into an alternative color-tunable single photon emission system with potentially broader spectral tunability compared to conventional color-tunable single photon emitters. We demonstrate that addition of an organic FA cation additive into all-inorganic CsPbBr$_3$ quantum dots can be used to fine-tune the emission wavelength across more than 30 nm in the visible, while retaining excellent single photon characteristics. We argue that the wide spectral tunability of color-tunable Cs$_{1-x}$FA$_x$PbBr$_3$ perovskite single photon emitters, independent of mechanical strain or applied fields, may facilitate their integration in quantum photonic systems.

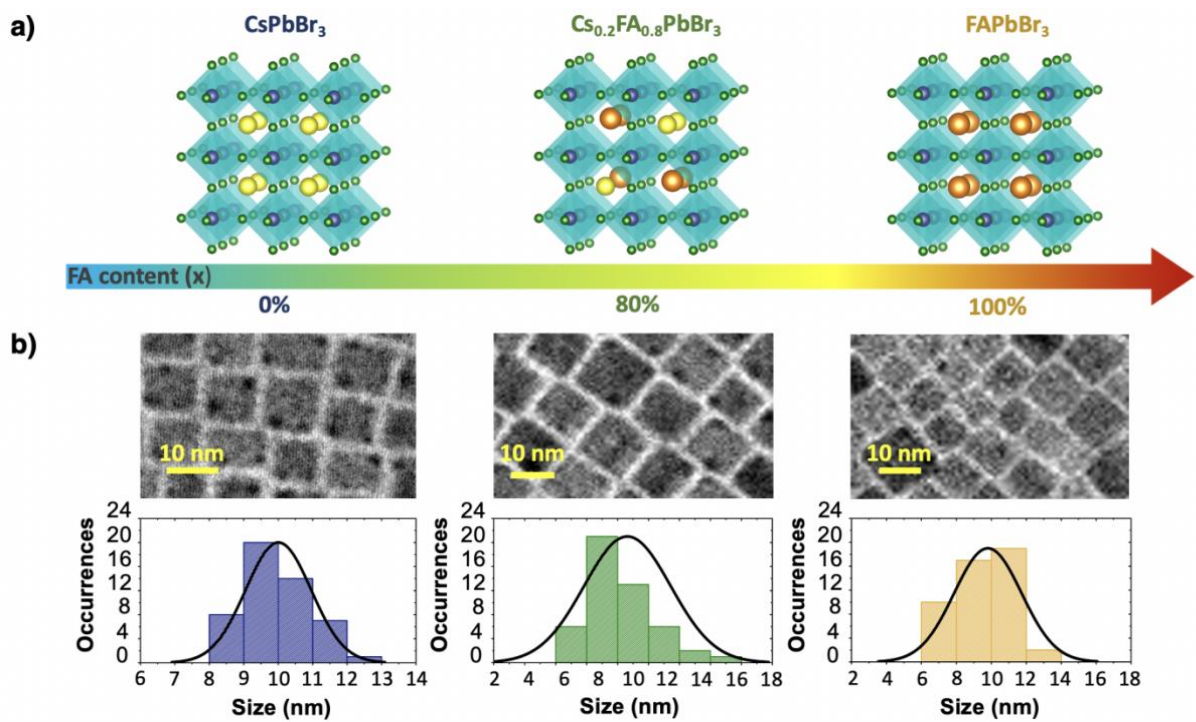

**Figure 1. Compositionally tunable Cs$_{1-x}$FA$_x$PbBr$_3$ quantum dots (x=0, 0.8 and 1).** (a) Changes in the crystallographic structures of the Cs$_{1-x}$FA$_x$PbBr$_3$ quantum dots with the addition of the organic FA cation additives. (b) Transmission electron microscopy images of the CsPbBr$_3$ (blue), Cs$_{0.2}$FA$_{0.8}$PbBr$_3$ (green), and FAPbBr$_3$ (yellow) quantum dots alongside with their respective size distribution histograms. A sampling size of 50 quantum dots was recorded for each composition.

**Results and discussion**

*Compositionally tunable Cs$_{1-x}$FA$_x$PbBr$_3$ perovskite quantum dots*

In this work, Cs$_{1-x}$FA$_x$PbBr$_3$ quantum dots were synthesized following a wet chemical approach (see **SI.1** of Supplementary Information). **Figure 1a** illustrates the substitution of the inorganic Cs cation present in the crystallographic structure of CsPbBr$_3$ with the addition of the organic FA cation additive. This substitution can be achieved by varying the molar mass of formamidinium bromide (FABr) and cesium bromide (CsBr) precursors during the synthesis. Changes in the crystallographic structure with the addition of the FA cations are observed in the x-ray diffraction measurements and high-resolution transmission electron microscopy (TEM) (See **Figure S1** and **S2** of Supplementary Information). The TEM images of the synthesized $Cs_{1-x}FA_xPbBr_3$ quantum dots (x=0, 0.8 and 1) are shown in **Figure 1b**. The quantum dots are monodispersed and cubic shaped with an average size of (10 ± 1) nm for x=0, (11 ± 3) nm for x=0.8, and (10 ± 2) nm for x=1 No significant alteration to the shape and size of synthesized quantum dots is observed when the Cs cations are substituted with the FA additives. This suggests that the quantum size effect and dimensionality will not play a substantial role in the alteration of their optical properties. Small size quantum dots with radius comparable to the excitonic Bohr radius (7 nm for $CsPbBr_3$[4] and 8 nm for $FAPbBr_3$[25]) are found in the respective size distribution histograms, indicating the presence of mixed-cation $Cs_{1-x}FA_xPbBr_3$ quantum dots which should be subject to quantum confinement effects and therefore feasible for single photon generation.

*Spectral tunability of $Cs_{1-x}FA_xPbBr_3$ quantum dot ensembles*

To investigate the effect of FA additives on the optical properties of the quantum dots, photoluminescence measurements were performed on $Cs_{1-x}FA_xPbBr_3$ quantum dot ensembles (x=0, 0.6, 0.8, and 1). **Figure 2a** shows the dependence of the PL spectra on addition of the FA cation additives. A red shift in the central emission wavelength (CEW), from 511 nm (x=0) to 523 nm (x=0.6), 532 nm (x=0.8), and 537 nm (x=1) can be clearly observed. The modification of the emission wavelength can be attributed to changes in the Pb–Br bond lengths and angles in the $PbBr_6^{4-}$ octahedron when Cs cations are substituted with FA. As the valence and conduction bands originate from the Br 4p and Pb 6p orbits, respectively, the modification of Pb–Br bond lengths and angles alters the electronic band structure of the quantum dots.[19] The full width half maximum (FWHM) of 19 nm (x=0 and 0.6), 20 nm (x=0.8), and 21 nm (x=1) is mainly due to inhomogeneous size distribution. These steady-state PL measurements demonstrate the possibility to control and fine-tune the emission wavelength of $Cs_{1-x}FA_xPbBr_3$ quantum dot ensembles via chemical engineering.

The PL kinetics of mixed-cation perovskite quantum dot ensembles is shown in **Figure 2b**. The time-resolved PL measurements also reveal a clear dependence of the average lifetime of the quantum dots on addition of the FA cation additives (**Figure 2c**, upper panel). The average lifetime was estimated to increase from 2.3 ns (x=0) to 15.3 ns (x=0.6), 42.0 ns (x=0.8) and 118.7 ns (x=1) by fitting the PL decays with a tri-exponential function (refer to the fitting parameters in **Table S1** of Supplementary Information). The increment in the lifetime with the increase of the FA/Cs ratio can be attributed to an increase in the excitonic binding energy.[19] Notably, the $Cs_{1-x}FA_xPbBr_3$ quantum dots show a high PLQY of 63–73%, regardless of their relative cation composition (**Figure 2c**, lower panel). Thus, addition of the FA cations does not cause a significant reduction of the quantum dot brightness.

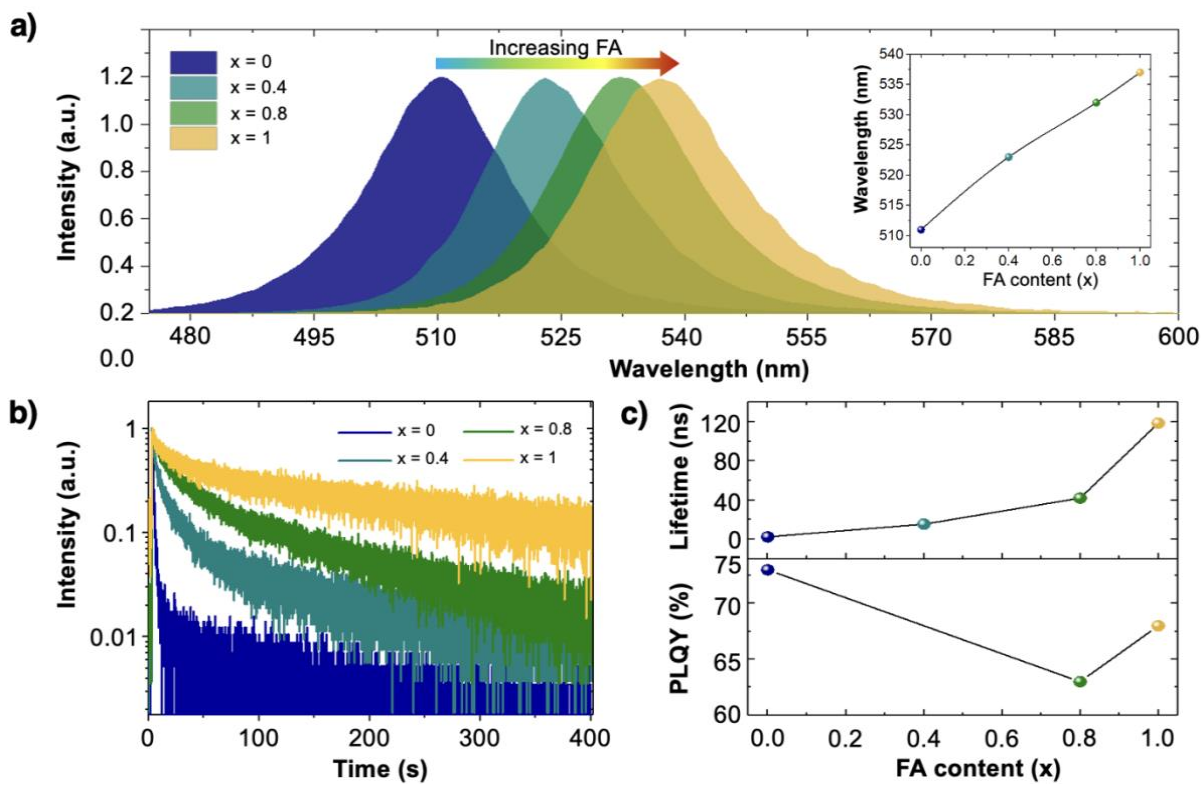

**Figure 2. Spectral tunability of the $Cs_{1-x}FA_xPbBr_3$ quantum dot ensembles (x=0, 0.4, 0.8, and 1).** (a) Normalized PL spectra of the $CsPbBr_3$ (blue), $Cs_{0.4}FA_{0.6}PbBr_3$ (cyan), $Cs_{0.2}FA_{0.8}PbBr_3$ (green), and $FAPbBr_3$ (yellow) quantum dot ensembles. The insert illustrates the shift in the CEW with the different amount of FA content. (b) Time-resolved PL lifetime measurements of the $Cs_{1-x}FA_xPbBr_3$ quantum dot. (c) Changes in the average lifetime (upper panel) and PLQY (lower panel) with the different amount of FA content.

*Spectral tunability of $Cs_{1-x}FA_xPbBr_3$ single photon emitters*

By using a confocal microscope, we were able to investigate the $Cs_{1-x}FA_xPbBr_3$ quantum dots at the single particle level. As the changes in the emission spectrum and lifetime upon addition of FA in quantum dot ensembles were attributed to intrinsic mechanisms such as the alteration of crystallographic bond lengths and angles, and the modification of the exciton recombination kinetics, compositional engineering is expected to bring about similar tunability to the optical properties of single quantum dots. **Figure 3a** displays the typical PL spectra of single CsPbBr$_3$, Cs$_{0.2}$FA$_{0.8}$PbBr$_3$ and FAPbBr$_3$ quantum dots, with CEW of 509 nm, 519 nm, and 527 nm, respectively. A red shift in the single dot PL spectra is observed upon substitution of Cs cations with FA, consistent with the trend observed in quantum dot ensembles. The spectral broadening of single quantum dot luminescence, in terms of FWHM, is 15 nm (x=0), 17 nm (x=0.8), and 19 nm (x=1). Likewise, an increment in the average lifetime from 5 to 36 ns is observed from typical PL decay kinetics of single CsPbBr$_3$, Cs$_{0.2}$FA$_{0.8}$PbBr$_3$, and FAPbBr$_3$ quantum dots (**Figure 3b** and fitting parameters in **Table S2** of Supplementary Information).

To achieve statistical significance, the optical properties of 20 Cs$_{1-x}$FA$_x$PbBr$_3$ quantum dot emitters were analyzed for each composition. **Figure 3c** shows the CEW distribution obtained for the three sets of Cs$_{1-x}$FA$_x$PbBr$_3$ quantum dots (x=0, 0.8 and 1). The average CEWs are 505.2 $\pm$ 3.4 nm for x=0, 517.1 $\pm$ 6.4 nm for x=0.8 and 523.5 $\pm$ 6.0 nm for x=1, where the deviation from average values is due to the inhomogeneous size distribution of the quantum dots (as also shown in Figure 1b). Thus, addition of FA additives allows extending the single photon emission wavelength towards the pure green, closing the single photon emission gap between pure CsPbBr$_3$ and FAPbBr$_3$ quantum dots.

**Figure 3d** shows the distribution of the FWHM of the PL spectra of individual Cs$_{1-x}$FA$_x$PbBr$_3$ quantum dots with representative compositions (x=0, 0.8 and 1) plotted against their CEW. The FWHM ranges from 14 to 16 nm for x=0, from 14 to 18 nm for x=0.8 and from 16 to 20.5 nm for x=1. By comparison with the PL linewidth of quantum dot ensembles, we attribute this slightly smaller FWHM to the fact that confocal measurements target small particles in the quantum confined regime. The clear correlation between the peak wavelength and the FWHM of single dot PL seen in Figure 3d points to the crucial role of exciton confinement to achieve narrowband emission.

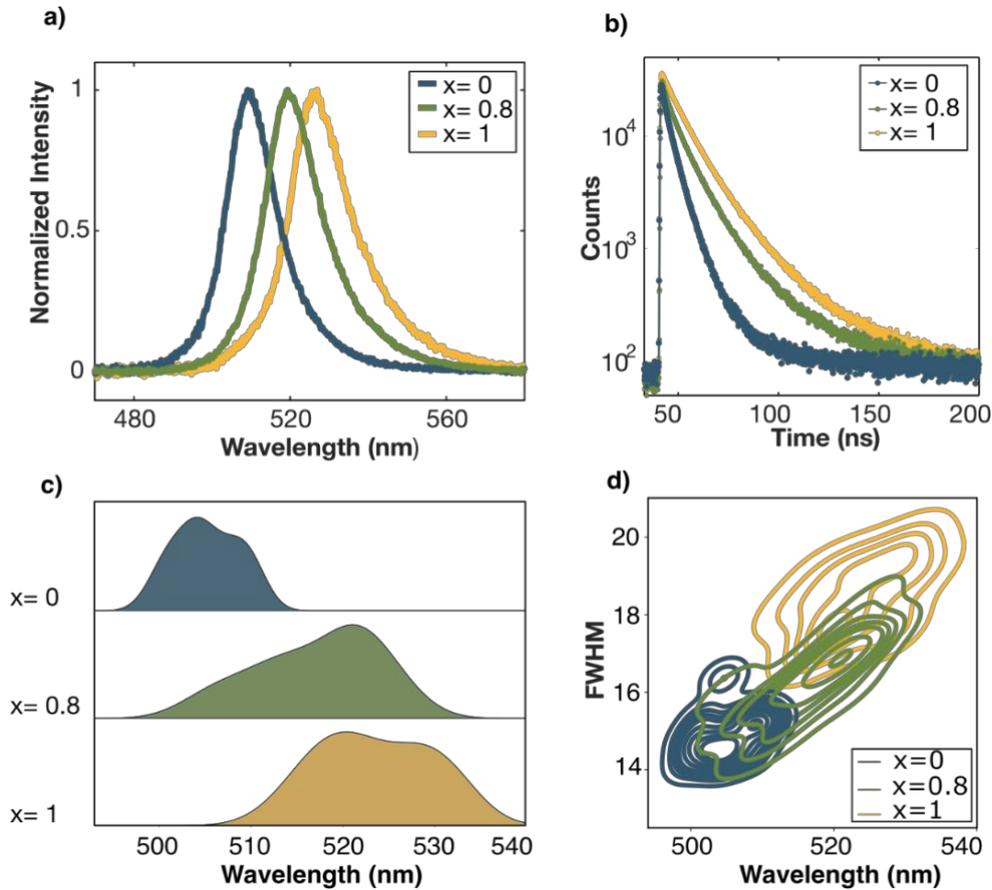

**Figure 3. Spectral tunability of the $Cs_{1-x}FA_xPbBr_3$ single photon emitters (x=0, 0.8 and 1).** a) Normalized emission spectra of an individual $CsPbBr_3$ (blue), $Cs_{0.2}FA_{0.8}PbBr_3$ (green), and $FAPbBr_3$ (yellow) quantum dot. The emission spectra are normalized to the peak intensity. b) Typical PL decay of a $CsPbBr_3$ (blue), $Cs_{0.2}FA_{0.8}PbBr_3$ (green) and $FAPbBr_3$ (yellow) quantum dots. c) Density distribution of the CEW for $CsPbBr_3$ (blue), $Cs_{0.2}FA_{0.8}PbBr_3$ (green), and $FAPbBr_3$ (yellow) quantum dots. A sampling size of 20 emitters for each composition was recorded. d) Distribution of the FWHM plotted against the relative central emission wavelengths for the $CsPbBr_3$ (blue), $Cs_{0.2}FA_{0.8}PbBr_3$ (green) and $FAPbBr_3$ (yellow) emitters.

*Photo-stability and blinking of $Cs_{1-x}FA_xPbBr_3$ single photon emitters*

In previous studies, the performance of perovskite quantum dot devices has often been hampered by their photo-instability under illumination. In particular, irreversible photon-induced degradation and PL intermittency between the high and low intensity states, also referred as photo-bleaching and blinking, are frequently reported in the literature of colloidal perovskites nanocrystals.[26-29] The photo-stability of $Cs_{1-x}FA_xPbBr_3$ single photon emitters (x=0, 0.8, and 1) was evaluated by measuring their PL time-traces, that is the intensity of emission as a function of time under steady-state excitation. The integration time and binning time were 600 s and 10 ms, respectively. The typical PL time-traces for individual x=0, 0.8 and 1 quantum dots are reported in **Figure 4 (panels a, c and e)**. All three compositions show intermittency between the high and low emissive states, and this is corroborated by the presence of two levels in the corresponding PL intensity histograms (shown on the right of the panels). The x=1 composition exhibits the most pronounced photo-blinking compared to x=0.8, and x=0 samples. Moreover, under continuous illumination, a significant decrease in PL intensity is observed, with greater reduction in the case of x=1, followed by the x=0.8, and x=0 perovskite quantum dots. The decrease in intensity over time is clearly manifested by the width of the PL intensity histograms, where the intensity peak is well-defined for x=0, becomes slightly broader in the case of x=0.8 and spreads widely for x=1 emitters. The observed photo-bleaching is a typical signature of photo-chemical degradation of perovskites QDs.[30] The relatively higher photo-instability demonstrated in the organic-inorganic FAPbBr$_3$ (x=1) perovskites with respect to the all-inorganic CsPbBr$_3$ (x=0) is well reported in literature.[31]

The duration of the low intensity (OFF) periods was evaluated by thresholding the PL time traces. We selected a threshold intensity of 15 counts/10 ms, considering as OFF states the intensities below this value. We then looked at the cumulative distribution of the OFF periods $P_{off}(\tau_{off} > \tau)$, which gives the probability for the emission to be OFF for a time period $\tau_{off}$ longer than a duration $\tau$. This probability distribution can be fitted by a power law with an exponential cut-off of the form[32]:

$$P_{off} = C\, t^{-m_{off}}\, e^{-t/\tau_c}, \qquad (1)$$

where $C$ is a constant, $t$ refers to the time, $m_{off}$ is the power low exponent and $\tau_c$ is the truncation cut-off. Fitting the cumulative distribution of the OFF periods for the Cs$_{1-x}$FA$_x$PbBr$_3$ QDs with different composition (see **Figure S9** of Supplementary Information), we obtain truncation cut-offs and power law exponents of $\tau_c$=0.25 s and $m_{off}$=1.34 (x=0), $\tau_c$=0.02 s and $m_{off}$=1.36 (x=0.8) and $\tau_c$=0.15 s and $m_{off}$=0.83 (x=1). For power law distributions with exponents smaller than 1, the so-called Levy's distributions, long OFF periods are very probable,[33] as in the case of x=1 composition. Conversely, an exponent greater than 1 is found when long OFF periods are very unlikely and the QD blinking is reduced,[34,35] as in the case of x=0.8 and, to a slightly lesser extent, x=0 composition. Here, we show that using the mixed-cation perovskites to tune the single photon emission can also reduce photo-instability, assuring reduced blinking and bleaching with respect to organic-inorganic FAPbBr$_3$ (x=1) perovskites.

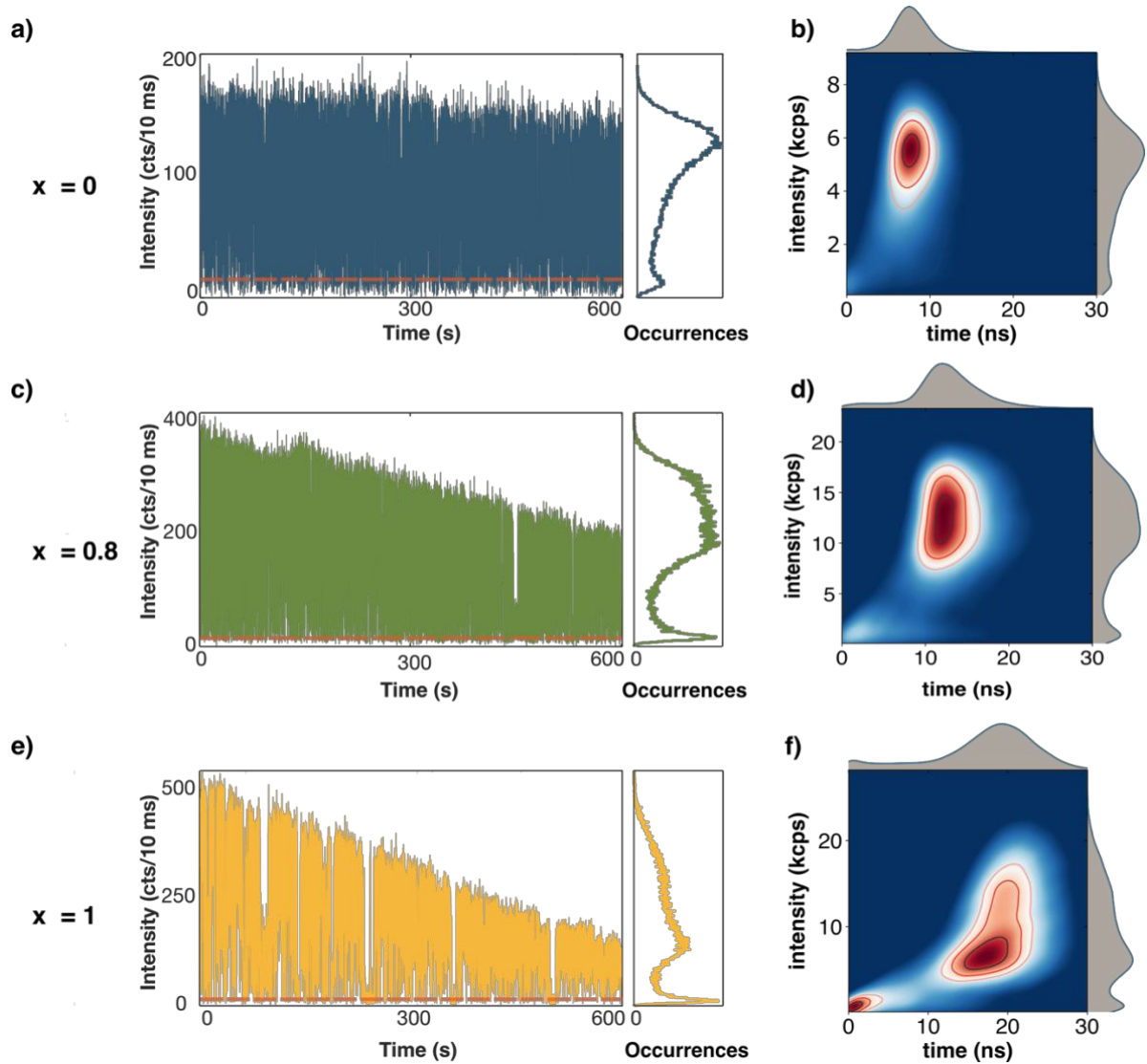

**Figure 4. Photo-stability and blinking of $Cs_{1-x}FA_xPbBr_3$ quantum dots (x=0, 0.8, and 1).** a), c), e) PL intensity time-traces and relative intensity histograms for single QDs with, respectively, a x=0, x=0.8 and x=1, with a binning time of 10 ms and an integration time of 600 s. The background level is given at 5 counts/10 ms. The threshold for the OFF states is 15 counts/10 ms (red line). b), d), f) False color representation of FLID images for NCS with, respectively, x=0, 0.8, and 1, obtained with a bin of 10 ms. A color change from blue to red corresponds to increasing probability of occurrences.

To evaluate the PL intermittency more quantitatively, we analyzed the correlations between PL intensity fluctuations and PL lifetime of individual quantum dots using the fluorescence lifetime-intensity distribution (FLID) analysis (for details on the analysis method refer to **SVII.5** in Supplementary Information). The correlation between these two quantities provides information on the origin of blinking in mixed-cation quantum dots. Indeed, the dependence of PL lifetime on the emitted intensity is a signature of type-A blinking, as opposed to type-B blinking[36] where PL intensity fluctuations occur without appreciable changes in lifetime. Type-A blinking fluctuations are known to be caused by a charging/discharging mechanism: while emission from the neutral exciton state is bright, the PL yield of the charged exciton state is reduced by the opening of non-radiative channels, like non-radiative Auger decay, which compete with the radiative recombination channel and quench photon emission. FLID images obtained using a bin of 10 ms for the three QD compositions are plotted in false color, changing from blue to red with increasing probability, in **Figure 4 (panels b, d and f)**. The resulting FLID trajectory is a curved line, typical of type-A blinking and confirming a correlation between PL intensity and lifetime. Moreover, for the x=0 and 0.8 emitters that showed reduced blinking the emission is found to be mostly in the bright state, while low intensity states became frequent for the x=1 emitters, and the effect of bleaching is more pronounced.

### *Single photon emission from $Cs_{1-x}FA_xPbBr_3$ quantum dots*

The $Cs_{1-x}FA_xPbBr_3$ quantum dots have an average size of 10 nm, with a minimum value of 7 nm (x=0) and 6 nm (x=0.8 and 1), as seen in Figure 1b. Given the excitonic Bohr radii of 7 nm (x=0) and 8 nm (x=1), quantum confinement and single photon emission, are to be expected from these quantum dots. Measurements of the second-order intensity correlation function $g^{(2)}(\tau)$ were conducted in a Hanbury Brown-Twiss configuration (**Figure S6**). The $g^{(2)}(\tau)$ function, which represents the probability to emit more than one photon per excitation pulse at a particular time $\tau$, is given by:

$$g^{(2)}(\tau) = \langle I(t)\, I(t+\tau) \rangle / \langle I(t) \rangle^2, \qquad (2)$$

where t is the time, $\tau$ the delay between two photon arrival event, and I (t) and I ($\tau$) the PL intensity at time t and $\tau$, respectively. At $\tau = 0$, $g^{(2)}(0)$ approaches zero when bi- and multi-excitonic radiative recombination are highly suppressed. **Figure 5a** shows the $g^{(2)}(0)$ values plotted in a semilogarithmic scale as a function of their CEW. A sample size of 20 emitters were measured for each of the three quantum dot compositions. We find that 97% of the emitters show $g^{(2)}(0)$ well below 0.5, a clear signature of single photon emission in all $Cs_{1-x}FA_xPbBr_3$ quantum dots. The majority of $CsPbBr_3$ (56%), $Cs_{0.2}FA_{0.8}PbBr_3$ (55%), and $FAPbBr_3$ (75%) quantum dots have $g^{(2)}(0)<0.1$. A strong antibunching behavior has been previously reported for all-inorganic perovskites[37] and $FAPbBr_3$[38] single photon emitters, and attributed to strong non-radiative Auger recombination, which effectively suppresses multiphoton emission events.[14] While 97% of the $Cs_{1-x}FA_xPbBr_3$ single photon emitters have $g^{(2)}(0)<0.5$, we also observe emitters with $g^{(2)}(0)$ values near and above 0.5. These particular emitters tend to have longer emission wavelength relative to other emitters of the same composition. As such, one could attribute the degradation and/or loss of single photon emission (quality of the antibunching) to the larger size quantum dots, whereby quantum confinement is no longer effective. The autocorrelation functions with the best anti-bunching behaviour recorded for each quantum dot composition are shown in **Figures 5b-d**, yielding $g^{(2)}(0)$ values of 0.05 (x=0), 0.04 (x=0.8), and 0.013 (x=1). Note that the delay peaks close to $\tau = 0$ are higher than 1 due to the presence of blinking;[39] at large delays, when the time scale becomes comparable to the typical blinking duration, the peaks tend to 1. Thus, the value of $g^{(2)}(\tau)$ was normalized by setting the mean height of the peaks at $\tau \approx 10$ ms to be 1.[40,41] The background counts arising from the dark counts of the APDs were also subtracted as described in Section **SVII.6** of the Supplementary Information.

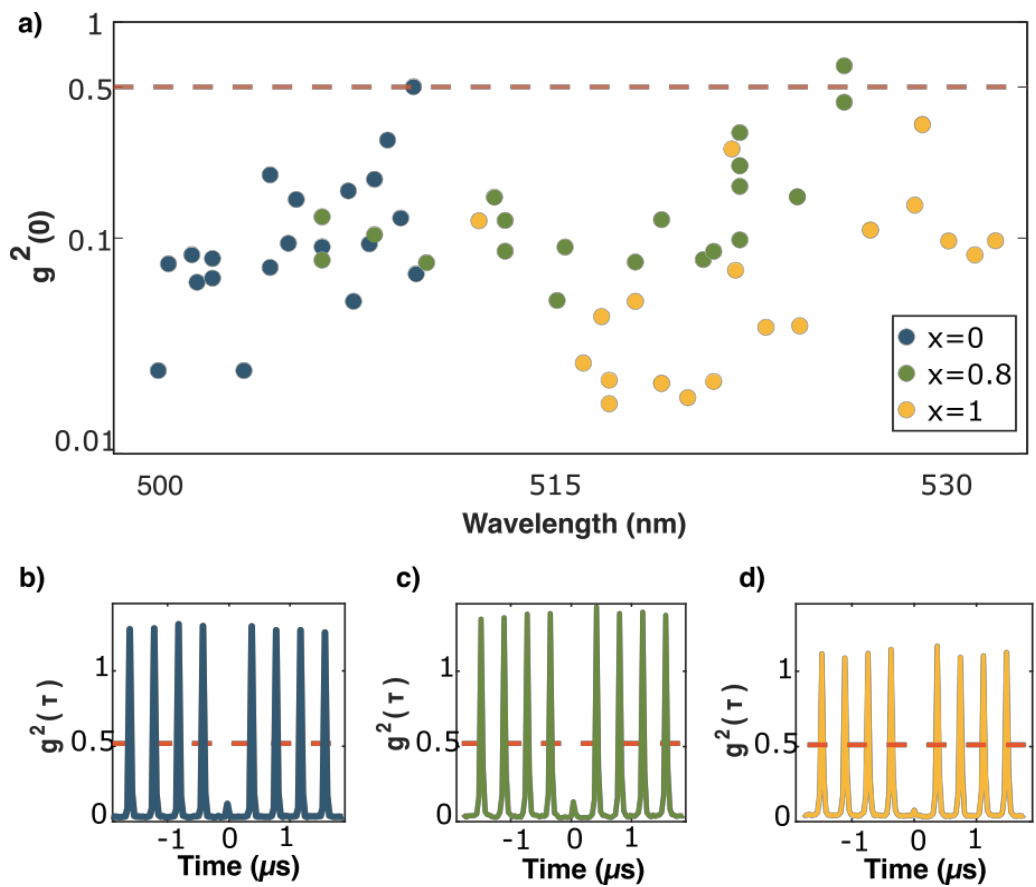

**Figure 5. Tunable single photon emission of the individual Cs$_{1-x}$FA$_x$PbBr$_3$ quantum dots (x=0, 0.8, and 1).** (a) The measured $g^{(2)}(0)$ values for CsPbBr$_3$ (blue), Cs$_{0.2}$FA$_{0.8}$PbBr$_3$ (green), and FAPbBr$_3$ (yellow) in the semi-logarithmic scale as a function of their CEW. A sampling size of 20 emitters per composition were taken and the threshold for antibunching is reported (red dashed line). (b) – (d) $g^{(2)}(0)$ function of an individual Cs$_{1-x}$FA$_x$PbBr$_3$ quantum dots (x=0, 0.8, and 1).

**Conclusion**

In summary, we have demonstrated that the color-tunable $Cs_{1-x}FA_xPbBr_3$ quantum dot system offers an alternative approach to fine-tune the single photon emission wavelength of perovskite quantum dots via chemical engineering. This approach could not only eliminate the need of applied mechanical strain or electric and magnetic fields to control the emission spectrum but offer colloidal quantum dots with excellent single photon characteristics and slightly broader spectral tunability (> 30 nm across the visible nm range) than conventional single photon emitters. We find that addition of the second cation additive closes the gap between $CsPbBr_3$ and $FAPbBr_3$ single photon emission, while improving stability. Notably, the strong antibunching of $Cs_{0.2}FA_{0.8}PbBr_3$ quantum dots indicates single photon emission quality comparable or even higher than room temperature single photon emitters like NV defect centres in diamonds[42] or III-V semiconductor quantum dots.[43] This opens new opportunities for integration of mixed-cation perovskite quantum dots into existing quantum photonic platforms.

**Acknowledgments**

We thank Maciej Klein for his help with the absorption and PLQY measurements of the QD ensembles, and Dong Shuyu for her assistance with X-ray diffraction measurements of the QDs. Research in NTU was supported by the Agency for Science, Technology and Research A*STAR-AME programmatic grant on Nanoantenna Spatial Light Modulators for Next-Gen Display Technologies (Grant no. A18A7b0058) and the Singapore Ministry of Education MOE Tier 3 (Grant no. MOE2016-T3-1-006). Research in LKB was supported by the ANR project IPER-Nano2 (ANR- 18CE30-0023) and by the European Union's Horizon 2020 research and innovation programme Nanobright (no. 828972). QG and AB are members of the Institut Universitaire de France (IUF).

**Data availability**

The authors declare that all data supporting the findings of this study are available within this article and its supplementary information and are openly available in NTU research data repository DR-NTU (Data) at https://doi.org/XXXXXX. Additional data related to this paper may be requested from the authors.

**Author contributions**

Q.Y.T. and C.S. conceived the idea. Q.Y.T. synthesized the perovskite QDs and characterized QD ensembles by TEM and PL measurements. Q.Y.T. measured absorption and PLQY of QD suspensions. M.D. prepared and optimized the deposition of low-density QD films, performed steady-state and time-resolved fluorescence spectroscopy at single particle level and $g^2(\tau)$ measurements with Time-Correlated Single Photon Counting (TCSPC) technique, statistical analysis of the optical properties, and characterized the blinking with Cumulative Distribution Function (CDF) and FLID analysis. Q.Y.T., M.D., A.B., and C.S. drafted the manuscript. All authors discussed the results and contributed to finalizing the manuscript. Q.G. and A.B supervised the research in LKB and C.S. supervised the research in NTU.

# Supplementary Information for

## Color-Tunable Mixed-Cation Perovskite Single Photon Emitters

Marianna D'Amato,[1†] Qi Ying Tan,[2,3†] Quentin Glorieux,[1]

Alberto Bramati,[1*] Cesare Soci[2,4*]

[1] Laboratoire Kastler Brossel, Sorbonne Universite, CNRS, ENS-PSL Research University, College de France, 4 place Jussieu, 75252 Paris Cedex 05, France

[2] Centre for Disruptive Photonic Technologies, The Photonics Institute, 21 Nanyang Link, Nanyang Technological University, Singapore 637371

[3] Interdisciplinary Graduate School, Energy Research Institute @NTU (ERI@N), Nanyang Technological University, 50 Nanyang Drive, Singapore 637553

[4] Division of Physics and Applied Physics, 21 Nanyang Link, School of Physical and Mathematical Sciences, Nanyang Technological University, Singapore 637371

*Correspondence to: csoci@ntu.edu.sg, alberto.bramati@lkb.umpc.fr

†These authors, alphabetically ordered, contributed equally to this work.

I. MATERIALS AND METHODS

1. Chemicals

Cesium carbonate ($Cs_2CO_3$, 99.9% trace metal basis), formaminidium bromide (FABr, ≥ 98%) lead (II) bromide ($PbBr_2$, 98%), oleylamine (OLA technical grade, 70%), N, N-Dimethylformamide (DMF, 99.8%), and toluene (99.8%) were purchased from Sigma-Aldrich. Octadecene (ODE, technical grade 90%), and oleic acid (OA, technical grade 90%) were purchased from Alfa Aesar.

2. Synthesis of $CsPbBr_3$ quantum dots

For the synthesis of Cs-oleate, 0.326 g of $Cs_2CO_3$, 18 mL of ODE and 1 mL of OA were loaded into a 100 mL three-neck flask. The mixture was heated under vacuum at 100°C for an hour and subsequently raised to 150°C under nitrogen flow. After complete dissolution, the solution was kept at 150°C to avoid solidification. For the synthesis of the $CsPbBr_3$ quantum dots, 0.067g of $PbBr_2$, 5 mL of ODE, 0.5 mL of OA and 1mL of OLA were loaded into a separate 100 mL three-neck flask. The mixture was heated under vacuum at 100°C for an hour and subsequently raised to 150°C under nitrogen flow. After complete dissolution, the solution was further heated up to 170°C. 0.6 mL of the as-prepared Cs-oleate was quickly injected to induce crystallization of the quantum dots. The solution was immediately placed into an ice – water bath and cooled to room temperature. To purify the resulting solution, ethyl acetate was subsequently added into the solution and the solution was centrifuged at 6000 rpm for 5 minutes. The precipitate was collected and re-dispersed into toluene. The solution was again centrifuged at 6000 rpm for 5 minutes and the supernatant was stored in a nitrogen filled environment to prevent degradation of the solution.

3. Synthesis of $Cs_{1-x}FA_xPbBr_3$ quantum dots

In a typical synthesis of $Cs_{0.2}FA_{0.8}PbBr_3$ quantum dots, 0.04 mmol of CsBr, 0.16 mmol of FABr, and 0.2 mmol of $PbBr_2$ were dissolved in 5 mL of DMF. After complete dissolution, 100 μL of OA and 50 μL of OLA were added into the solution. 200 μL of the precursor solution was injected into 5 mL of toluene under stirring to induce crystallisation of the quantum dots. The solution was left stirring for 24 hours before storage. All synthesis was conducted at room temperature and pressure. Quantum dots of different Cs and FA cation composition were achieved by varying the stoichiometric amount of CsBr, FABr, and $PbBr_2$.

4. **Characterization methods**

*Transmission electron microscopy (TEM)* was performed on a JEM-1400 flash electron microscope operating at 100 kV. The TEM samples were prepared by drop casting the solution onto a 400 mesh copper grids with carbon supporting films.

*X-ray diffraction measurements* were performed on a Rigaku SmartLab X-ray diffractor. Spin-coated samples on a glass microscope substrate were prepared for the measurements.

*UV-Vis absorbance measurements* of the $Cs_{1-x}FA_xPbBr_3$ quantum dots solution were performed on an UV-2600i UV-Vis spectrophotometer.

*Photoluminescence quantum yield (PLQY) measurements* of the $Cs_{1-x}FA_xPbBr_3$ quantum dots solution were performed on a integrating sphere with an excitation laser of $\lambda=435$ nm.

*Photoluminescence (PL) measurements* were performed on an upright microscope set-up. A fibre coupled picosecond pulsed laser (Pico Quant P-C-405B) with an output wavelength $\lambda=405$ nm and repetition rate of 2.5 MHz was used as an excitation source. The laser beam was focused onto the sample with a 10X microscope objective lens. The emitted PL was collected back by the same objective, filtered with a long pass filter (cut off wavelength of $\lambda=420$ nm), and detected by a spectrometer (Acton SpectroPro 2300i). For time-resolved PL measurements, the collected emission is detected with a single photo avalanche detector (Micro Photon Devices) connected to a single photon counting module (Pico Harp 300). All PL measurements were carried out at room temperature.

## III. X-RAY DIFFRACTION OF $Cs_{1-x}FA_xPbBr_3$ QUANTUM DOTS

Figure S1 shows the X-ray diffraction (XRD) patterns of the $Cs_{1-x}FA_xPbBr_3$ quantum dots. The diffraction pattern of the $CsPbBr_3$ quantum dots compares well with those of the cubic $CsPbBr_3$ phase (JDCDS No.75-0412). With an increasing FA cation content, a shift in the diffraction peaks towards a lower angle is observed (Figure S1, insert). Such an observation suggests the expansion of the d-spacing due to the incorporation of the larger radius FA cations.

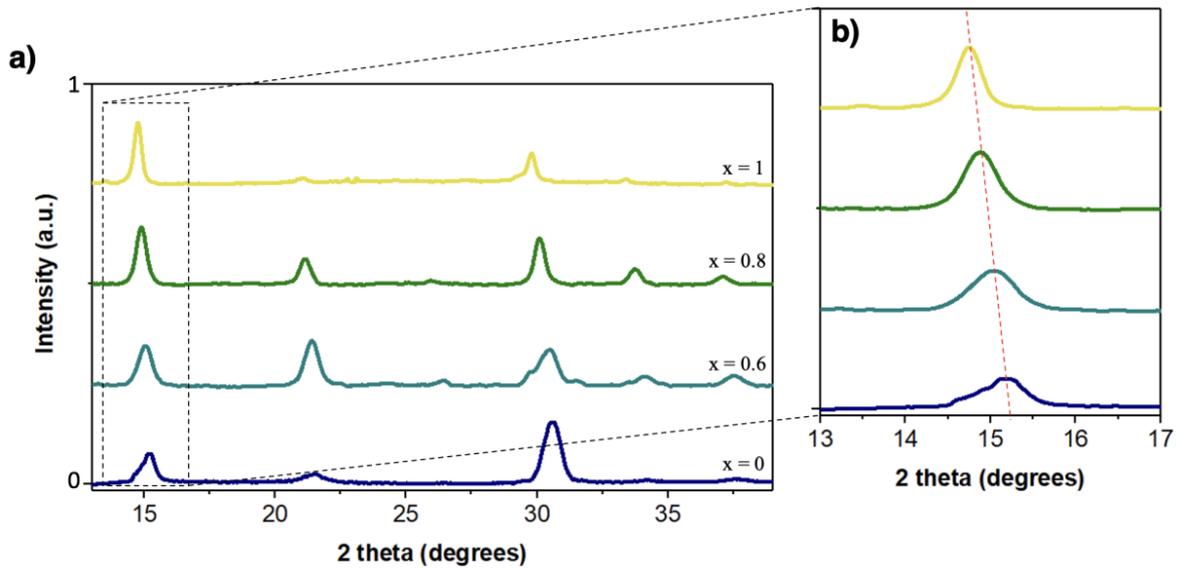

**Figure S1. XRD patterns of the Cs$_{1-x}$FA$_x$PbBr$_3$ quantum dots.** (a) XRD pattern of x=0,0.5,0.8, and 1). Diffraction pattern of the CsPbBr$_3$ quantum dots corresponds to the cubic CsPbBr$_3$ phase. (b) Zoom of the diffraction peaks around 15°. The gradual shift in the peaks (towards a lower angle is attributed to the substitution of a smaller radius Cs cation with a larger radius FA cation.

**IV. HIGH-RESOLUTION MICROSCOPY OF Cs$_{1-x}$FA$_x$PbBr$_3$ QUANTUM DOTS**

Figure S2 shows the high-resolution transmission electron microscope (HRTEM) images of the Cs$_{1-x}$FA$_x$PbBr$_3$ quantum dots. The observed increment in the d-spacing from 4.1 Å (x=0), 5.3 Å (x=0.8), to 6.0 Å (x=1) coincides with the findings from the XRD measurements. Furthermore, HRTEM microscopy reveals single-crystalline nature with high crystallinity. Similar to the TEM observations, no significant changes in the shape and size of the quantum dots is observed with the addition of FA cations.

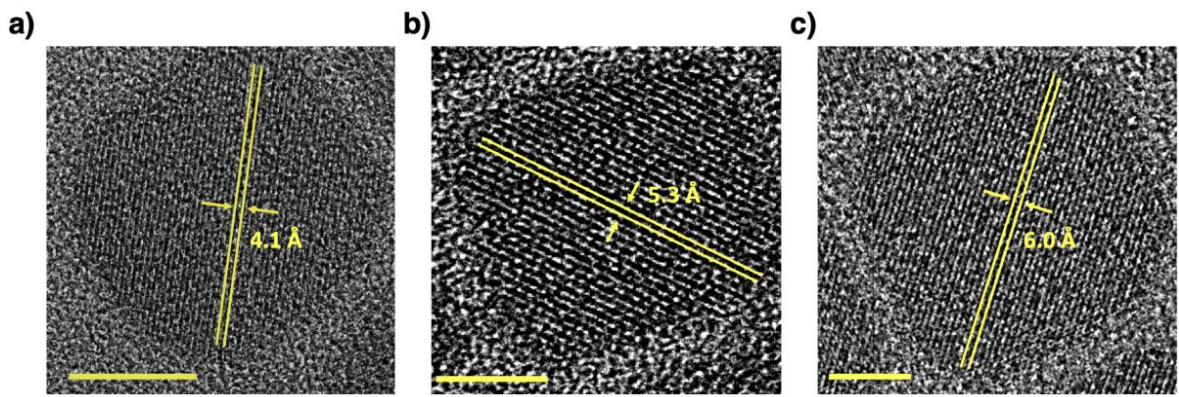

**Figure S2. HRTEM images of the Cs$_{1-x}$FA$_x$PbBr$_3$ quantum dots.** (a) x=0, (b) x=0.8, and (c) x=1. The scale bars represent 5 nm.

## IV. ABOSRBPTION AND PHOTOLUMINESCENCE OF $Cs_{1-x}FA_xPbBr_3$ QUANTUM DOT ENSEMBLES

Figure S3 shows the absorbance and photoluminescence spectra of the $Cs_{1-x}FA_xPbBr_3$ quantum dots. With the addition of the FA cations, a red-shift in the absorbance and photoluminescence spectra is observed, without any significant alteration in their FWHM. Such phenomena arises from the modification in the electronic band structure of the quantum dots.

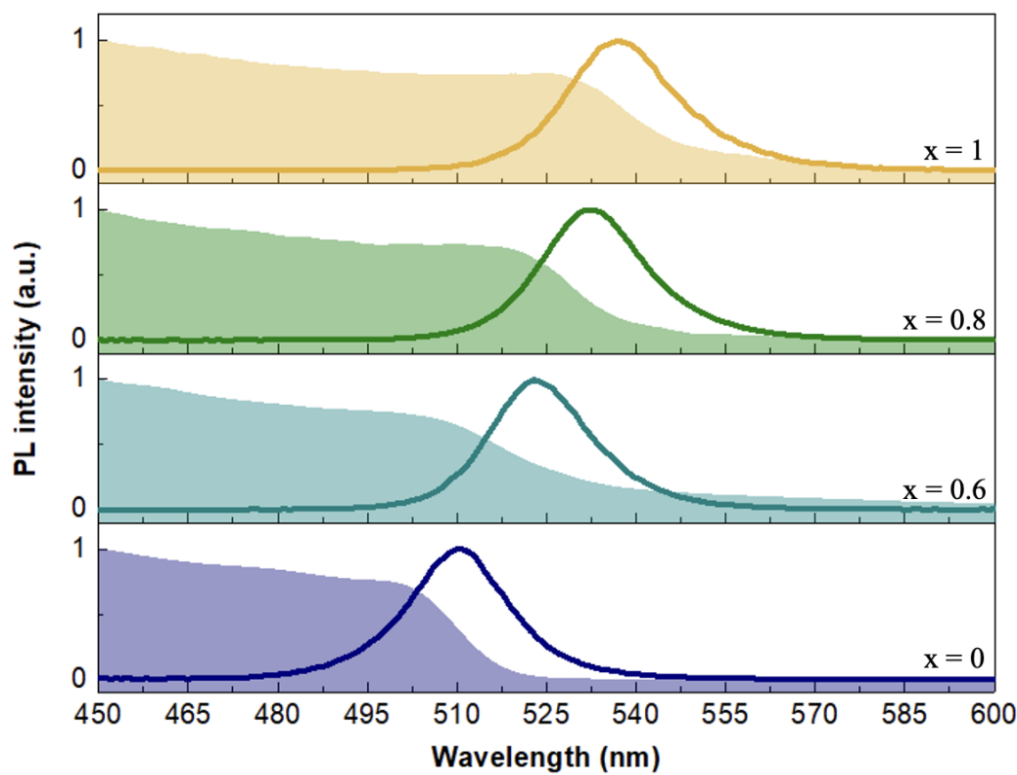

**Figure S3. Absorption and photoluminescence spectra of the $Cs_{1-x}FA_xPbBr_3$ quantum dots.** A red-shift in both the absorption and photoluminescence spectra of x=0,0.6,0.8, and 1 is observed and attributed to the changes in electronic band structure.

## V. STABILITY OF $Cs_{1-x}FA_xPbBr_3$ QUANTUM DOT ENSEMBLES

Figure S4 shows the photo-stability of the $Cs_{1-x}FA_xPbBr_3$ quantum dots. Photoluminescence spectra of the quantum dots were taken over the days. Changes in the CEW of each composition is analysed. A slight shift in the CEW is observed in the span of 9 days. The blue-shift observed of the $Cs_{0.4}FA_{0.6}PbBr_3$, $Cs_{0.2}FA_{0.8}PbBr_3$ and $FAPbBr_3$ quantum dots is attributed to the dissociation of the FA cations while the red-shift of the $CsPbBr_3$ quantum dots is attributed to the increment in the size.

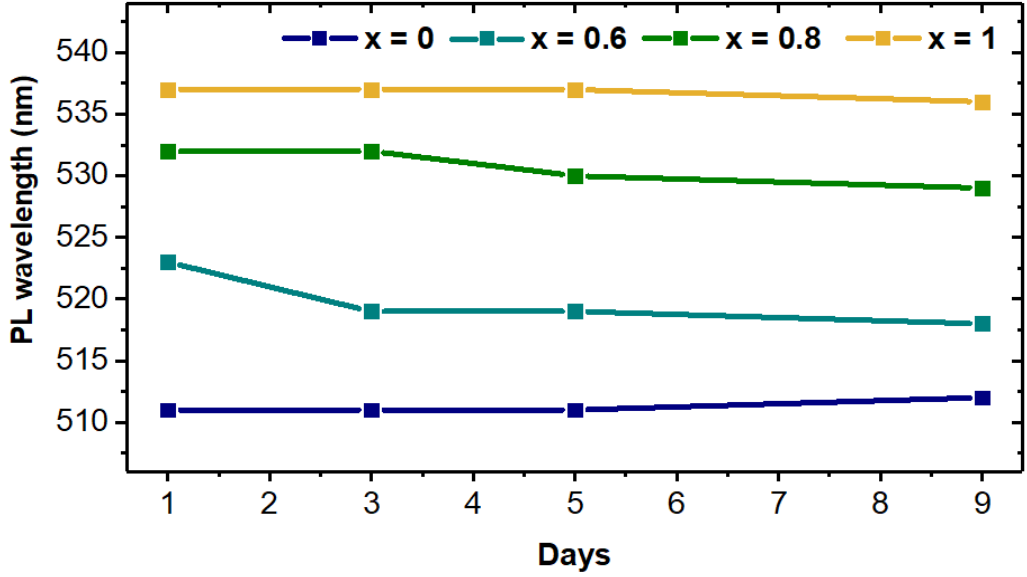

**Figure S4. Shift in CEW of the photoluminescence spectra of the $Cs_{1-x}FA_xPbBr_3$ quantum dots.** Changes in the CEW are attributed to the dissociation of the FA cations (x=0.6, 0.8 and 1) and changes in the size of the quantum dots (x=0).

## VI. TIME-RESOLVED LUMINESCENCE ANALYSIS

The PL decays of the quantum dot ensembles are fitted with a tri-exponential function as follows

$$A(t) = A_1 \exp\left(-\frac{x-x'}{\tau_1}\right) + A_2 \exp\left(-\frac{x-x'}{\tau_2}\right) + A_3 \exp\left(-\frac{x-x'}{\tau_3}\right) + B \quad (S1)$$

In Eq. S1, $\tau_i$ refers to the different lifetimes, $A_i$ are the amplitudes for each decay component, and B takes into the account the noise.

The fitting parameters used in Figure S5 are as listed in Table S1.

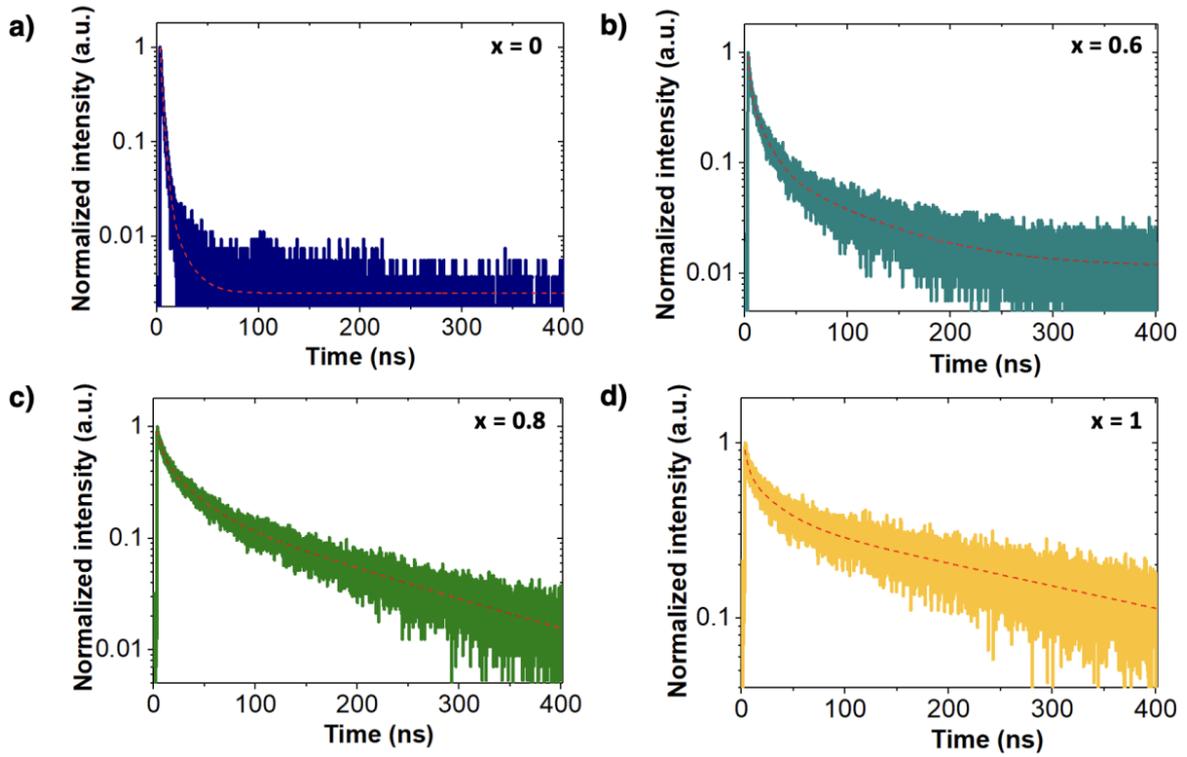

**Figure S5. Typical PL decays and fittings of the $Cs_{1-x}FA_xPbBr_3$ quantum dot ensembles.** The experimental results are modelled with a tri-exponential decay function (red dotted lines).

Table S1. Fitting parameters of the PL decays.

|  | x=0 | x=0.6 | x=0.8 | x=1 |
|---|---|---|---|---|
| $B$ | 0.003 | 0.01 | 0.002 | 0.001 |
| $A_3$ | 0.024 | 0.089 | 0.195 | 0.366 |
| $A_1$ | 0.46 | 0.489 | 0.346 | 0.235 |
| $A_2$ | 0.50 | 0.336 | 0.360 | 0.304 |
| $\tau_2$ | 1.35 | 13.1 | 28.3 | 30.0 |
| $\tau_1$ | 3.0 | 2.18 | 5.8 | 4.1 |
| $x'$ | 3.17 | 3.13 | 3.13 | 3.13 |
| $\tau_3$ | 13 | 79.0 | 150.0 | 337.0 |

## VII. SINGLE PARTICLE SPECTROSCOPY

### 1. Experimental set-up

Figure S6 shows the schematics of the inverted microscope set-up used for both wide-field and confocal microscopy in the single particle measurements. The $Cs_{1-x}FA_xPbBr_3$ quantum dots were diluted and spin-coated onto a fused-silica substrate. The surface density of the quantum dots was kept below 0.1 $\mu m^{-2}$ to spatially isolate the quantum dots. In the wide-field scheme, the substrate placed on the inverted microscope (Nikon Eclipse Ti) was illuminated with a LED lamp (CooLED pE-100) at λ = 400 nm and imaged with a CMOS camera (Hamamatsu ORCA-Flash 4.0) to locate the emitters. Upon locating the emitters, a picosecond pulsed laser (Pico Quant P-C-405B) with an output wavelength λ = 405 nm, pulse width < 50 ps and a repetition rate f = 2.5 MHz was used to excite the emitters. In this confocal scheme, the laser beam was focused onto the sample with a 100X oil immersion microscope objective with numerical aperture of 1.4. The emitted PL was collected back with the same objective, spectrally filtered with both a dichroic mirror and long pass filter (cut off wavelength of λ = 435 nm) to remove the excitation laser. The PL emission can be sent, interchangeably, to the CMOS camera, to a spectrometer and to a Hanbury Brown-Twiss set-up for time-correlated single photon counting detection. All measurements were conducted at room temperature.

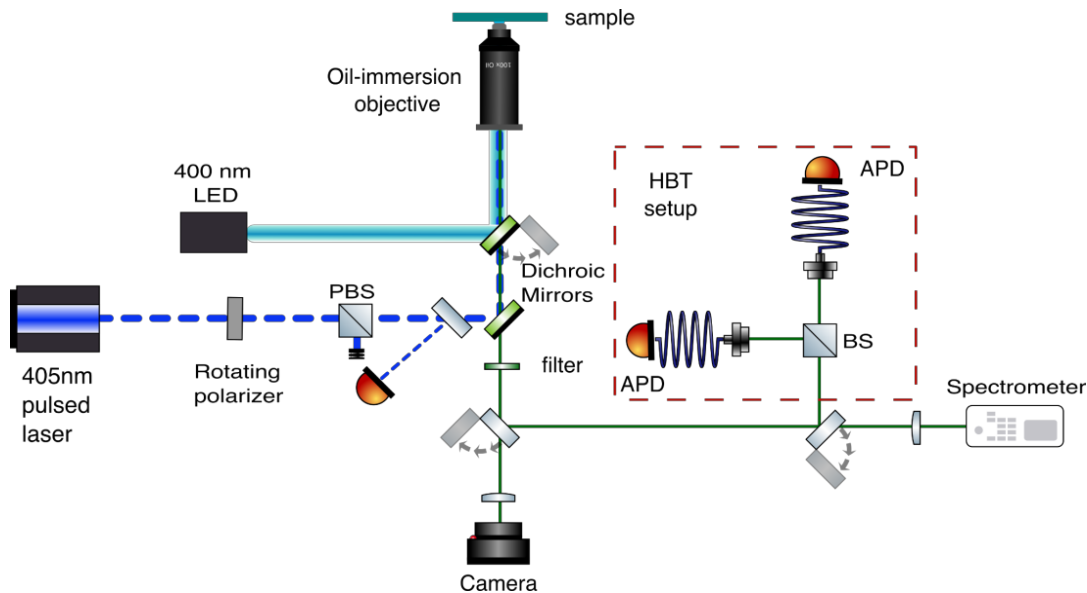

**Figure S6. Single particle level characterization.** Schematics of the inverted confocal microscope set-up used in the measurements. The set-up consists of both a wide-field microscope for imaging of the samples and confocal microscopy for single particle level characterization.

## 2. Photoluminescence intensity of a $Cs_{1-x}FA_xPbBr_3$ as a function of the excitation power

As a preliminary measure, for each nanocrystal the emitted intensity as a function of the excitation power is measured. It is well-known that for high confined colloidal nanocrystal at room temperature, the efficiency of the non-radiative Auger recombination process is strongly enhanced, becomes efficient and it is also responsible for the emitted intensity saturation usually observed. Indeed, when the quantum yields for the m-multiexciton state QYm (i.e. the radiative recombination probability of m e-h pairs in the nanocrystal per laser pulse) are suppressed by efficient Auger recombination, the probability of emitting more than one photon for excitation pulse is negligible and a perfect saturation curve is expected. When multi-exciton states are not quenched by any non-radiative relaxation process the trend of the PL is linear with the excitation power. In the intermediate case of multiexciton QYs not negligible, the contribution from the multi-excitonic emission modifies the shape of the saturation curve[S1]. It is possible to demonstrate that in this case the second order correlation function $g^2(0)$ will depend on the intensity at which we excite the emitters. To obtain then a reference to perform the characterization on several emitters, we performed all the measurements at the saturation power $P_{sat}$.

Figure S7 plots the PL intensity of a typical single $Cs_{1-x}FA_xPbBr_3$ emitter as a function of the excitation power. This excitation power was varied using a half-wave plate and the collected PL emission was recorded by the CMOS camera. The experimental data of the saturation curved is fitted using the following function

$$I = A\left[1 - e^{-\frac{P}{P_{sat}}}\right] + B\frac{P}{P_{sat}} \tag{S2}$$

where A and B refer to the exciton and biexciton contribution respectively. $P_{sat}$ is extracted from the fitting and used for single particle spectroscopy measurements.

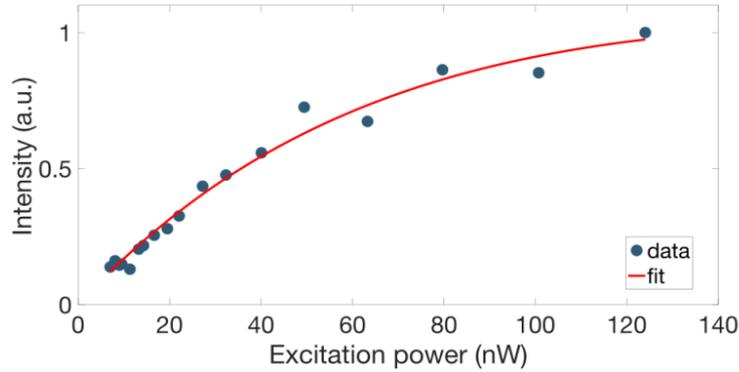

**Figure S7. Saturation measurement.** A typical PL saturation curve of a single QD. The experimental data (blue dots) are fitted with Eq. S2 (red line).

### 3. Photoluminescence decays

PL decays are fitted with a tri-exponential decay model:

$$I = A_1 e^{-(t-t_0)/\tau_1} + A_2 e^{-(t-t_0)/\tau_2} + A_2 e^{-(t-t_0)/\tau_3} + C \tag{S3}$$

where $\tau_1$, $\tau_2$, and $\tau_3$ are the fitted lifetimes, $t_0$ represents the pulse arrival time and $A_i$ are the amplitudes of each decay component; C is an constant to take into account the dark counts. In Figure S8 (panel a, b, c) shows the fitting of the PL decays. The fitting parameters are reported in table S2.

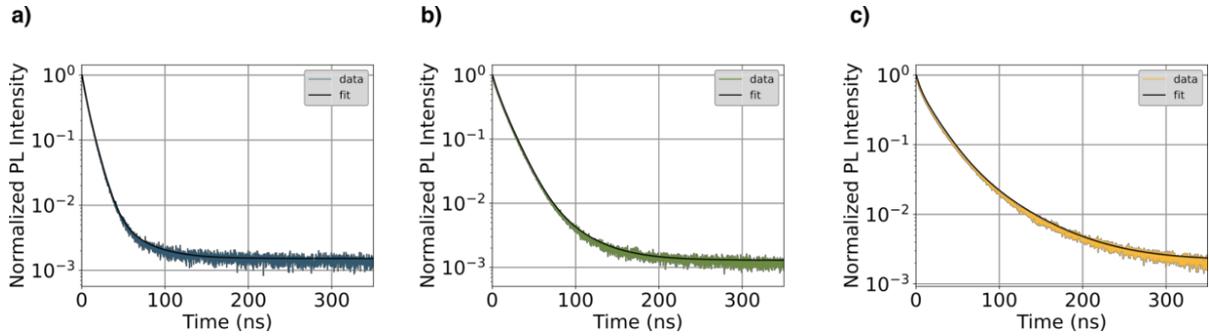

**Figure S8. Photoluminescence decays.** a-c) PL decays of single $Cs_{1-x}FA_xPbBr_3$ QDs, respectively with x=0 (blue), x=0.8 (green) and x=1 (yellow), fitted with Eq. S3 (black line).

**Table S2. Fitted lifetimes.** Fitting parameters of the PL decays of $Cs_{1-x}FA_xPbBr_3$ quantum dot, respectively for x=0, x=0.8 and x=1.

|  | $\tau_1$ | $A_1$ | $\tau_2$ | $A_2$ | $\tau_3$ | $A_3$ |
|---|---|---|---|---|---|---|
| x = 0 | 3.5 | 0.430 | 8.4 | 0.678 | 31.6 | 0.013 |
| x = 0.8 | 4.5 | 0.259 | 13.9 | 0.761 | 37.8 | 0.035 |
| x = 1 | 3.9 | 0.320 | 19.6 | 0.634 | 54.9 | 0.100 |

## 4. Cumulative distribution of the OFF period

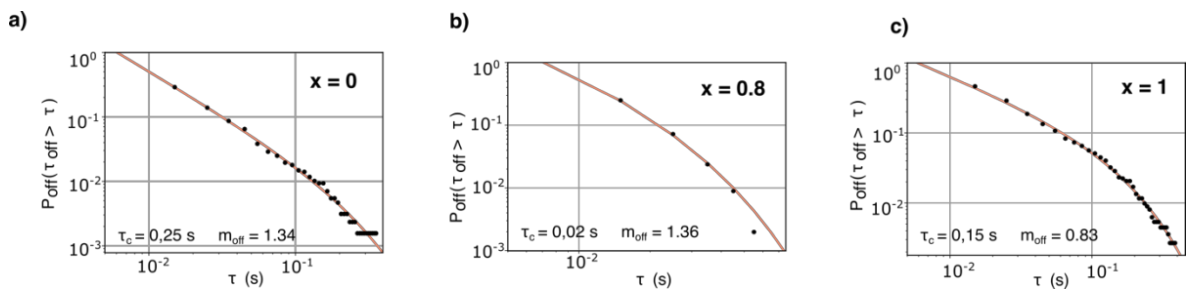

**Figure S9. Cumulative distributions of OFF periods.** Cumulative distribution of the OFF periods for the $Cs_{1-x}FA_xPbBr_3$ with, respectively, (a) x=0, (b) 0.8, and (c) 1. Data are fitted with Eq. 1 (in the main text).

## 5. Generation of the Fluorescence Lifetime Intensity Distribution (FLID) images

To obtain the FLID images in Figure 4, the kernel density estimation is used[S2,S3]. We started from the PL time-traces obtained with a binning time of 10 ms. For the photons detected in each bin of the PL intensity time-trace, the main arrival time is calculated, obtaining a corresponding lifetime time-trace . The upper panel of Figure S10 (panel a, b, c) show the PL intensity time-trace of the three QD compositions within an enlarged period of a few seconds, while the bottom panels show the corresponding lifetime trajectories. The main arrival time together with the bin's mean intensity, constitutes a point in the FLID intensity-time space.

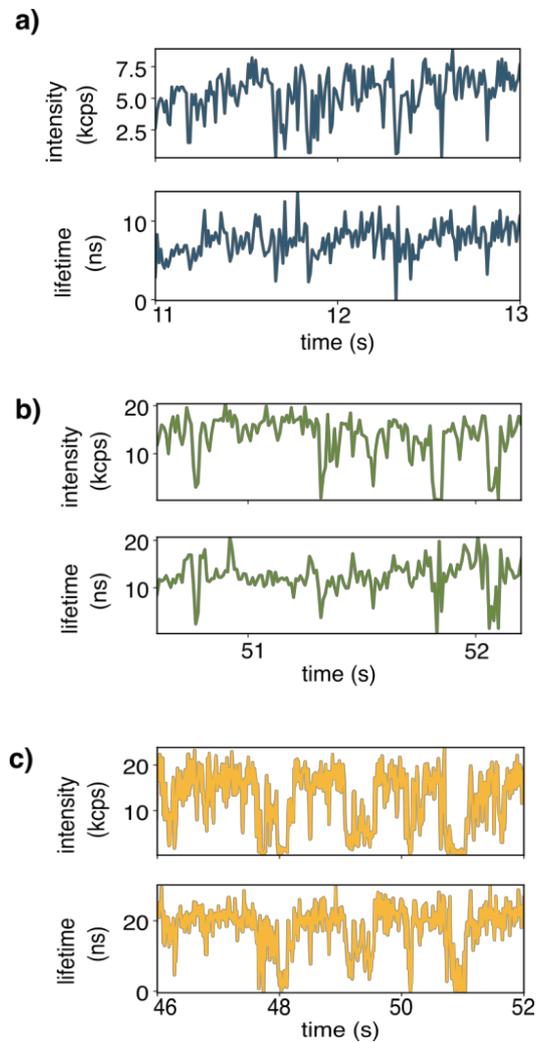

**Figure S10. PL and lifetime trajectories.** Zoom of PL intensity (upper panel) and lifetime (lower panel) trajectories for single NCs with respectively (a) x=0, (b) 0.8, and (c) 1.

## 6. Noise cleaning of g²(0) measurements

To clean the g²(τ) histogram from the background counts we start to consider the number of counts measured at a given delay τ, that we call S(τ). Each of these counts can be generated either by a start (or stop) from the signal, with a probability a(t), or by a start (or stop) from the background, with a probability b(t). We can write

$$S(\tau) = C\,(a(\tau) + b(\tau))\,(a(\tau) + b(\tau)) \qquad (S4)$$

where C is a constant of proportionality.

Now we consider a time $\tau_b$ in between two consecutive peaks, where there is no signal and then $a(\tau_b) = 0$. Calling $S' = \frac{S}{C}$, we obtain:

$$S'(\tau_b) = b^2(\tau) \qquad (S5)$$

Solving the system of two equations (2) and (3), we obtain the formula:

$$S_{clean}(\tau) = C\,a^2(\tau) = S(\tau) + S(\tau_b) - 2\sqrt{S(\tau)}\sqrt{S(\tau_b)} \qquad (S6)$$

with which we can clean the g²(0) measurements from the background counts.